\documentclass[11pt,twoside]{article}
\usepackage{./asp2014}

\aspSuppressVolSlug
\resetcounters

\begin{document}

\title{The yellow hypergiant -- B[e] supergiant connection}
\author{Anna Aret,$^1$ Michaela Kraus,$^{1,2}$ Indrek Kolka,$^1$ and Grigoris Maravelias$^2$
\affil{$^1$Tartu Observatory, 61602 T{\~o}ravere, Tartumaa, Estonia; \email{aret@to.ee}\\
$^2$Astronomick\'y \'ustav AV\v{C}R, v.v.i., Ond\v{r}ejov, Czech Republic}}

\paperauthor{Anna Aret}{aret@to.ee}{}{Tartu Observatory}{}{T{\~o}ravere}{Tartumaa}{61602}{Estonia}
\paperauthor{Michaela Kraus}{kraus@to.ee}{}{Tartu Observatory}{}{T{\~o}ravere}{Tartumaa}{61602}{Estonia}
\paperauthor{Indrek Kolka}{indrek@to.ee}{}{Tartu Observatory}{}{T{\~o}ravere}{Tartumaa}{61602}{Estonia}
\paperauthor{Grigoris Maravelias}{maravelias@asu.cas.cz}{}{Astronomick\'y \'ustav AV\v{C}R}{}{Ond\v{r}ejov}{}{25165}{Czech Republic}

\begin{abstract}
B[e] supergiants and yellow hypergiants share a number of common properties regarding their circumstellar environments. 
Using the forbidden [O\,{\sc i}] and [Ca\,{\sc ii}] lines  as disk tracers,
we suggest the presence of a
Keplerian disk or ring around the yellow hypergiant V509\,Cas and confirm the pole-on inner disk around V1302\,Aql. 
These findings indicate a change in mass-loss behavior from spherical in cooler yellow hypergiants to axisymmetric in the hotter ones during
the passage through the Yellow Void. The accumulation of material in the equatorial plane 
reminds of the disks of B[e] supergiants, supporting the suggestion that yellow hypergiants might appear as 
B[e] supergiants after they reach the blue edge of the yellow instability domain.
\end{abstract}

\section{Introduction}\label{introduction}
During their post-main-sequence evolution massive stars pass through several short-lived phases
of enhanced mass loss. Yellow hypergiants (YHGs) and B[e] supergiants (B[e]SGs) are well-known groups 
in such transition phases. The ejected material typically accumulates in either shells,
rings, or disk-like structures. 
Rich emission line spectra of such objects provide information on the structure and kinematics of
the circumstellar matter. Forbidden emission lines of singly ionized or neutral metals, 
such as  [Ca\,{\sc ii}] and [O\,{\sc i}], serve as particularly valuable tracers for their formation regions,
carrying information on kinematics and physical conditions.
It has been shown \citep*{2012MNRAS.423..284A,2016MNRAS.456.1424A} that the appearance of these lines
requires high-density environments combined with large emitting volumes. Such conditions are met in the innermost regions of
Keplerian rotating disks around B[e]SGs, where [Ca\,{\sc ii}] lines are formed closest to the star, 
followed by the [O\,{\sc i}] $\lambda$5577 line and the [O\,{\sc i}] $\lambda\lambda$6300, 6364 lines further out.

YHGs are massive stars that have passed through the red-supergiant (RSG) phase
and evolve back bluewards in the Hertzsprung-Russell diagram. Enhanced mass loss and eruptions in YHGs are ascribed to 
increased pulsation activity \citep[see the overview by][]{1998A&ARv...8..145D}. Although physical mechanisms causing build-up of
massive disks around B[e]SGs are unknown, it has been suggested that pulsations might play a role as well \citep{2016A&A...Kraus}. 
Controversial results have been obtained concerning the evolutionary state of B[e]SGs. While the enrichment in $^{13}$C in their disks
favors a pre-RSG state \citep{2013A&A...558A..17O}, their massive dusty environments indicate a post-RSG
evolutionary phase. Similarities in the circumstellar environments of YHGs and B[e]SGs support the latter case 
and the suggestion by \citet*{2007ApJ...671.2059D} that YHGs may be evolving toward the B[e]SG phase.

\section{Observations}\label{observations}

During the years 2010--2015, we carried out an optical spectroscopic survey of 
a large sample of Ga\-lac\-tic northern emission-line stars.
The survey includes four YHGs: V1302\,Aql (A2-F8\,Ia), V509\,Cas (A7-G5\,Ia), $\rho$\,Cas (F0-G7\,Ia), V1427\,Aql (F3-G5\,Ia),
and two B[e]SGs: V1478\,Cyg (B0-B1.5\,I) and 3\,Pup (A2.7\,Ib). The B[e]SGs were discussed in detail in \citet{2016MNRAS.456.1424A}.

Spectra were obtained using the Coud\'{e} spectrograph attached to the 
Perek 2-m telescope at Ond\v{r}ejov Observatory \citep{2002PAICz..90....1S} in
three wavelength regions:
around H~$\alpha$ (6250--6760\,\AA, $R\simeq$ 13\,000),
in the region of the [Ca\,{\sc ii}] $\lambda\lambda$7291,~7324 lines
(6990--7500\,\AA, $R\simeq$ 15\,000), and in the region of the
Ca\,{\sc ii} infrared triplet (8470--8980\,\AA, $R\simeq$ 18\,000).
The H~$\alpha$ region also encloses the two [O\,{\sc i}]
$\lambda\lambda$6300, 6364 lines.

\articlefigure{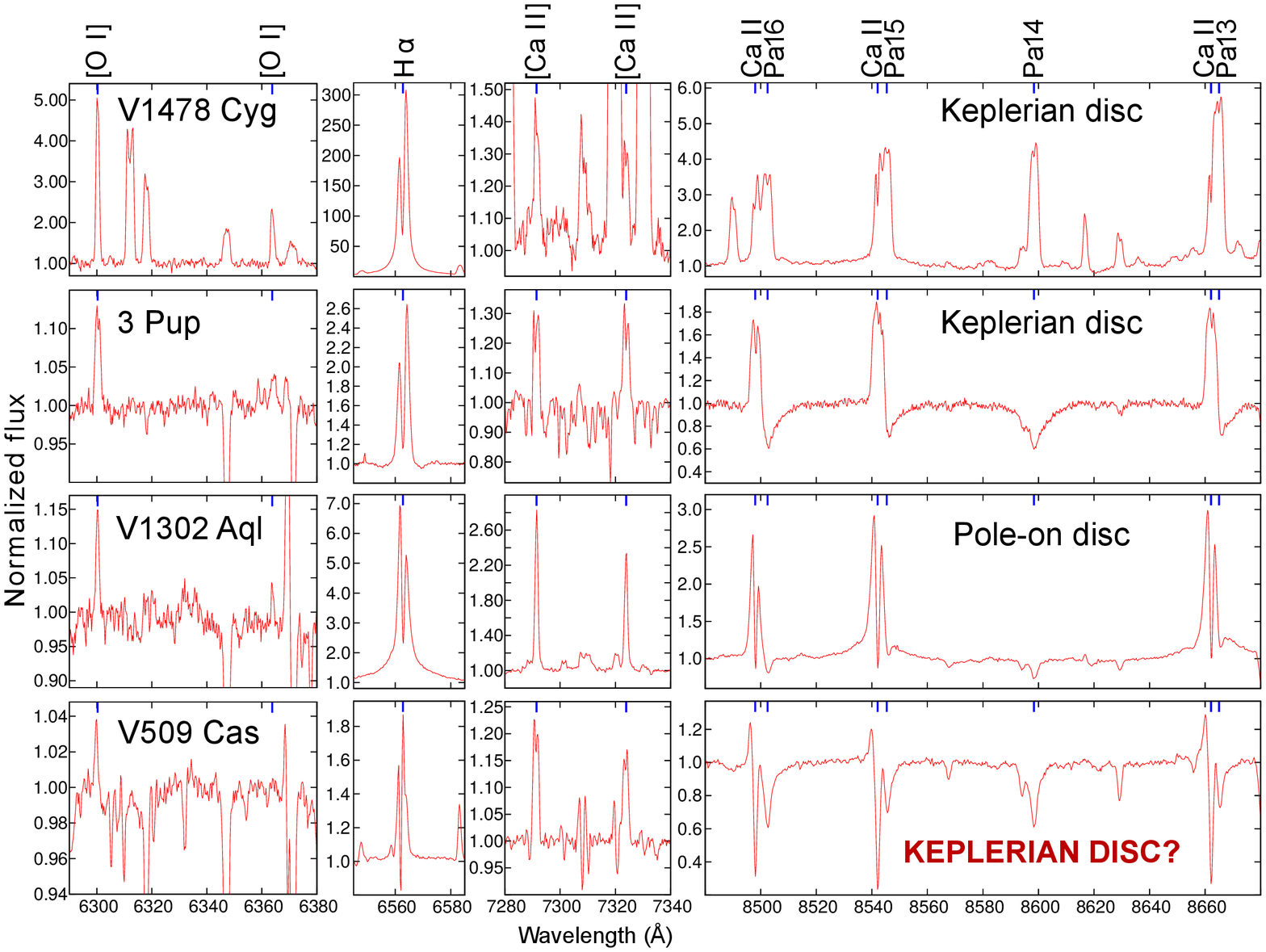}{fig1}{Spectra of the B[e]SGs 
compared to the two hotter YHGs.}

\section{Results and Conclusions}\label{results}

Both the [Ca\,{\sc ii}] and [O\,{\sc i}] lines were identified in the spectra of the B[e]SGs 
V1478\,Cyg and 3\,Pup, and of the hotter YHGs V1302\,Aql and V509\,Cas (Fig.\,\ref{fig1}), while 
in the cooler YHGs $\rho$\,Cas and V1427\,Aql
only the [Ca\,{\sc ii}] lines were present, but not the [O\,{\sc i}] lines. 
Also the infrared Ca\,{\sc ii} triplet lines are in emission in B[e]SGs and the hotter YHGs, 
while in the cooler
YHGs we see them in absorption. 
This indicates a difference in excitation mechanisms of the [Ca\,{\sc ii}] lines
in hotter and cooler environments.

\articlefigure{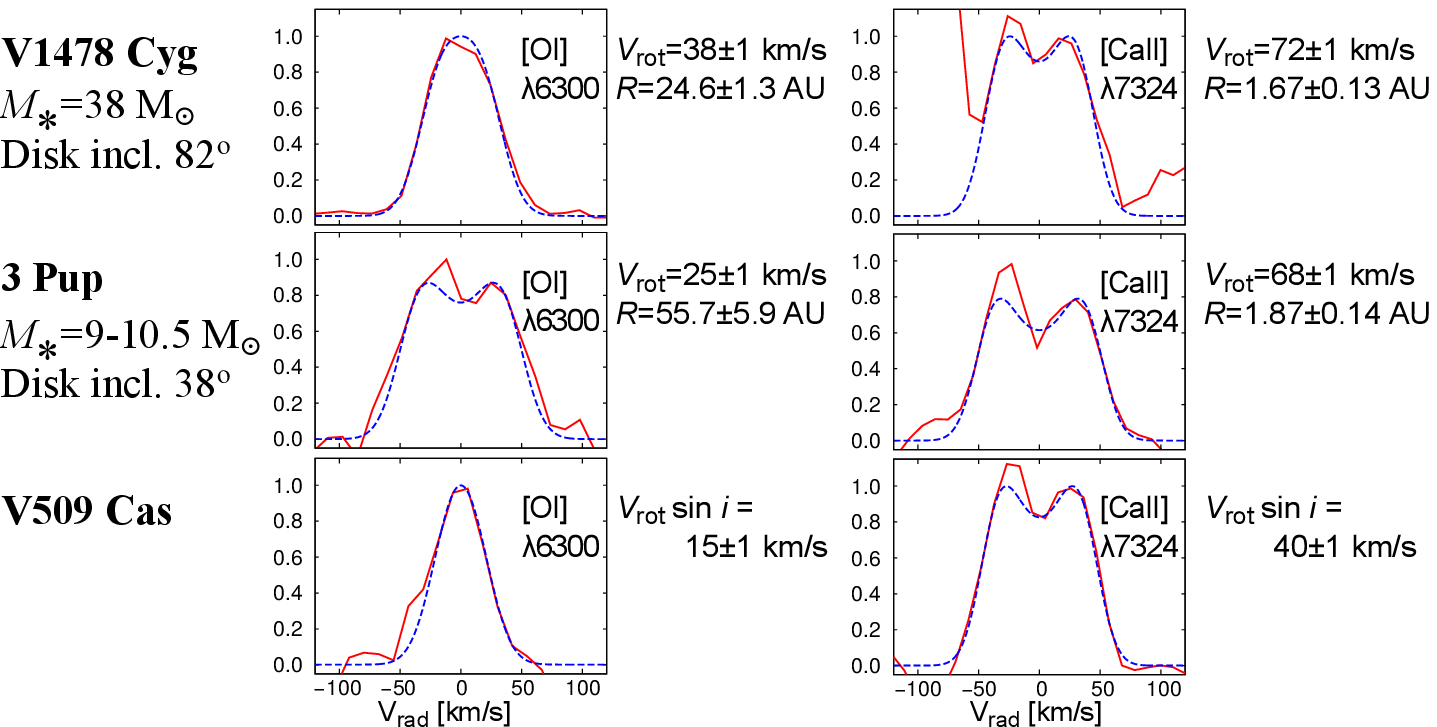}{fig2}{Kinematic model fits (dashed blue) to the observed profiles (solid red) of the forbidden lines. 
Shown are the obtained rotational velocities
and the corresponding radii of the emitting rings assuming Keplerian rotation.}

Both B[e]SGs are known to possess a disk \citep[see][]{2016MNRAS.456.1424A}.
We obtained rotational velocities from the line profiles of these B[e]SGs using a simple kinematic model, 
assuming that the emission originates from a narrow Keplerian rotating ring. 
We calculated the profile shape considering only the 
rotational velocity projected to the line of sight and the resolution of the spectrograph (Fig.\,\ref{fig2}).

The presence of both sets of forbidden lines in the two hot YHGs indicates that 
the physical conditions in their environments could be similar to those in the B[e]SGs. 
The profiles of the forbidden emission lines of V1302\,Aql are  
single peaked (Fig.\,\ref{fig1}) and display no kinematical broadening beyond spectral resolution. 
Such profiles are in agreement with their formation in the nearly pole-on seen disk \citep{2007A&A...465..457C,2010AJ....140..339T}.
Not much is currently known concerning the structure of the small-scale environment of V509\,Cas.
The forbidden lines in our spectra  show clearly 
double-peaked profiles for the [Ca\,{\sc ii}] lines. The [O\,{\sc i}] 
lines appear as broad but single peaked features at our spectral resolution,
however,  modeling of the profile reveals significant rotational broadening of the lines, as for V1478\,Cyg (Fig.\,\ref{fig2}).
This kinematical picture is typical for B[e]SGs \citep{2010A&A...517A..30K, 2012MNRAS.423..284A,2016arXiv161000607M}. 
Thus, V509\,Cas is the second YHG with clear indication for an inner disk.

In both cooler YHGs the profiles of the [Ca\,{\sc ii}] lines are very narrow single-peaked.
They provide no information on the kinematics in their [Ca\,{\sc ii}] line-forming regions,
but are consistent with an origin from a spherically symmetric shell.

\articlefigure{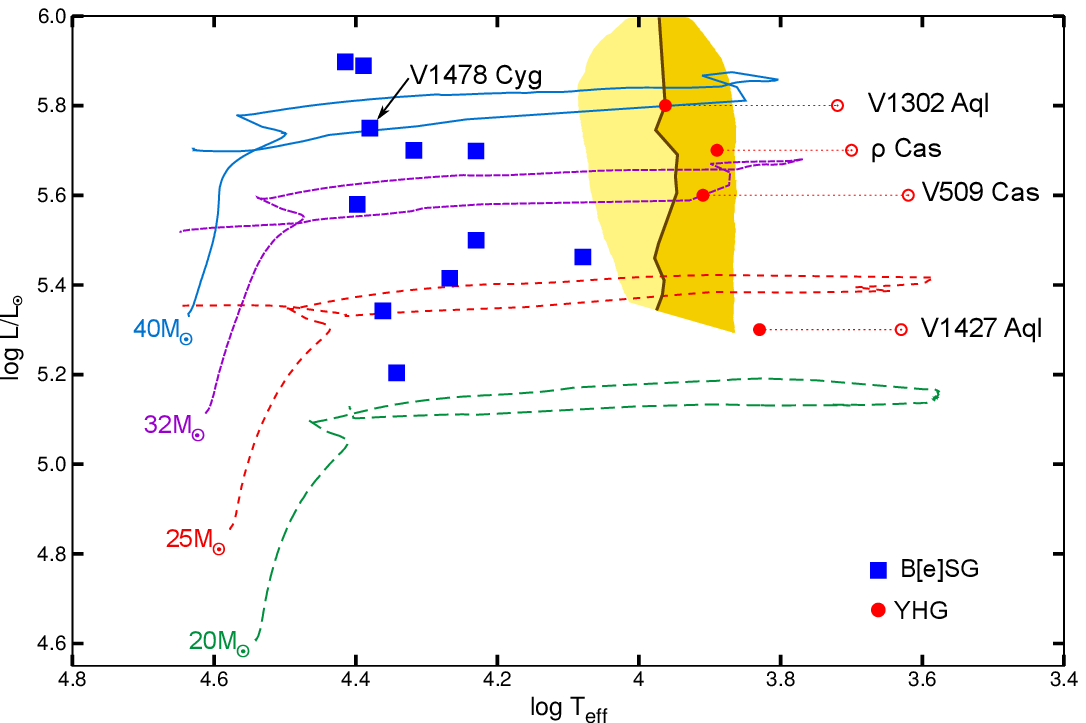}{fig3}{Upper part of the HRD
illustrating a possible evolutionary link between YHGs and
B[e]SGs. For the YHGs horizontal excursions due to
$T_{\rm eff}$ variations are shown. Evolutionary tracks are from
\citet{2012A&A...537A.146E}. The Yellow Void is as in
\citet{1997MNRAS.290L..50D}, the "first" instability region is darker and its
high-temperature boundary is marked by the solid line \citep{2012A&A...546A.105N}.}
 
The presence of disks around the hotter YHGs might indicate a drastic change in
the mass-loss behavior from spherical to axisymmetric during the evolution through 
the Yellow Void, as was  suggested by \citet{2007ApJ...671.2059D}. As pulsations are believed to be the 
main mechanism responsible for enhanced mass loss and eruptions in YHGs 
\citep{1998A&ARv...8..145D}, this would require a change 
in the pulsation habit of the stars when they approach the blue boundary of the "first" instability region (Fig.\,\ref{fig3}).  
During their passage through the 
second part of the Yellow Void, more material might still accumulate in the equatorial plane and
at the blue edge of the Yellow Void the stars might appear as B[e]SGs. 
Such a  scenario was already suggested for V1302\,Aql by \citet{Zickgraf1998} and 
\citet{2007ApJ...671.2059D}. It might also apply to V509\,Cas and even to a larger sample 
of hot YHGs, which is worth being investigated in more detail. 

\acknowledgements A.A. and I.K. acknowledge financial support from the Estonian 
grant IUT40-1; M.K. and G.M. from GA\,\v{C}R (14-21373S) and  RVO:67985815.

\end{document}